\documentclass[prl,reprint,twocolumn,showpacs,superscriptaddress]{revtex4-2}
\usepackage[export]{adjustbox}
\usepackage{amsmath,amssymb,bm,mathrsfs,graphicx, braket, times, mathtools}
\usepackage[colorlinks=true,citecolor=blue,linkcolor=blue]{hyperref}
\usepackage{multirow}
\usepackage[all,cmtip]{xy}
\usepackage[normalem]{ulem}
\usepackage[usenames,dvipsnames]{color}
\usepackage{tikz-cd}
\usepackage[compat=1.0.0]{tikz-feynman}
\usepackage{bbding}
\usepackage{graphicx}
\usepackage{subfigure}

\usepackage[utf8]{inputenc}
\usepackage[english]{babel}
\usepackage{amsthm}
\usepackage{wasysym}

\begin{document}

\title{Quantum metric driven transition between superfluid and incoherent fluid}

\author{Xuzhe Ying} \email{yingxz@ust.hk}
\author{Kangle Li}
\affiliation{Department of Physics, Hong Kong University of Science and Technology, Clear Water Bay, Hong Kong, China}

\begin{abstract}
We study the interplay between repulsive interaction and superfluidity in flat band system. We consider spatially indirect excitons in the Lieb lattice bilayer as an example. We show that due to the presence of repulsive interaction, the excitons may form an incoherent fluid. By increasing the quantum metric, the exciton fluid experiences a transition into the superfluid phase. Such transition may be captured by a model of Josephson junction array, which is beyond mean field description.
\end{abstract}

\maketitle

The study of flat band systems (including twisted multilayer graphene \cite{cao2018correlated,cao2018unconventional,sharpe2019emergent,serlin2020intrinsic,cao2021nematicity,xie2021fractional,tian2023evidence} and transition metal dichalcogenide \cite{cai2023signatures,PhysRevLett.132.036501,gu2022dipolar,ma2021strongly,xu2020correlated,park2023observation}) puts the concept of quantum geometric tensor to the central stage \cite{provost1980riemannian,PhysRevB.56.12847,cheng2010quantum,PhysRevResearch.2.023237,PhysRevB.90.165139,PhysRevB.96.165150,PhysRevLett.127.170404,PhysRevLett.132.026002,mao2023diamagnetic,kruchkov2023spectral,onishi2024quantum,PhysRevB.105.L241102,PhysRevB.62.1666,PhysRevB.98.220511,PhysRevB.101.060505,PhysRevB.106.014518,PhysRevLett.132.236001,verma2024instantaneous,doi:10.1073/pnas.2106744118,PhysRevB.105.L140506,rossi2021quantum,komissarov2024quantum,PhysRevB.109.024507}. Given that the kinetic energy is quenched in flat band systems, the wavefunction as well as the quantum geometric information encoded within wavefunctions becomes important. It was realized that the nontrivial quantum geometric properties makes various correlated phases of matter possible in flat bands, including fractional Chern insulator \cite{PhysRevResearch.2.023237,PhysRevB.90.165139,PhysRevB.96.165150}, superconductivity (SC) or superfluidity (SF) of excitons \cite{PhysRevLett.127.170404,PhysRevLett.132.026002,tian2023evidence,PhysRevB.101.060505,PhysRevB.106.014518,PhysRevLett.132.236001,doi:10.1073/pnas.2106744118,PhysRevB.105.L140506}. In the rest of the manuscript, our discussion will be limited to the superfluid of excitons \cite{eisenstein2004bose,doi:10.1146/annurev-conmatphys-031113-133832,PhysRevLett.68.1379,PhysRevLett.68.1383,nandi2012exciton}.

In flat band systems, the mean field studies neatly demonstrate the role of quantum metric in establishing superfluid through the quantity of superfluid stiffness \cite{PhysRevLett.127.170404,PhysRevLett.132.026002,tian2023evidence,PhysRevB.101.060505,PhysRevB.106.014518,PhysRevLett.132.236001,doi:10.1073/pnas.2106744118,PhysRevB.105.L140506}. Nevertheless, quantum fluctuations are important in flat band system \cite{PhysRevLett.130.226001}. Strong competing quantum fluctuation could potentially destroy superfluidity. One source of quantum fluctuation comes from the repulsive interaction between excitons. The repulsive interaction has a tendency to localize the excitons and thus desctroys the global phase coherence. Conventionally, the competition between repulsive interaction and superfluidity amounts to the comparison between typical repulsive interaction energy and the kinetic energy. A quantum phase transition between Wigner crystal and superfluid were observed numerically \cite{PhysRevB.84.075130,filinov2009effective,PhysRevB.76.064511}. In quantum Hall bilayer, similar phenomena were observed experimentally \cite{liu2022crossover,zeng2023evidence}.

In this manuscript, we address the question of how repulsive interaction and superfluidity compete with each other in \emph{flat band} systems. In flat band systems, we show that quantum metric or electronic wavefunction is very important. We found that the SF fluid is more stable in a regime with larger quantum metric or larger exciton density. In the opposite limit, repulsive interaction introduces strong quantum fluctuation, which destroys the global phase coherence of SF, even though a nonzero mean field order parameter is still possible. The state without global phase coherence is an incoherent fluid. We point out that the transition between superfluid and incoherent fluid states can be solely driven  by a change in the electronic wavefunction, or quantum metric.

Technically, we employ the model of Josephson junction array to describe the competition between repulsive interaction and the superfluidity. In the regime with large quantum metric, the Josephson coupling is generally stronger, in which case the superfluid phase is favored. In the opposite limit, global phase coherence is easily destroyed by the repulsive interaction, leading to an incoherent fluid consist of incoherent excitons. Certain technical details can be found in the Supplemental Material (SM) \cite{SupplementalM}.

We emphasize that our treatment based on Josephon junction array is really beyond mean field description. Weak competing fluctuation, as introduced by weak repulsive interaction, presumably does not kill the mean field order parameter. As we will argue that a nonzero mean field order parameter does not necessarily imply superfluidity. One has to study the dynamics of low energy excitations. Even though qualitative expectation can be drawn from the study of superfluid stiffness, the continuous description of superfluid stiffness is not enough to capture the competition between repulsive interaction and superfluid. This is reason for studying a Josephson junction array model on lattice. Indeed, the model of Josephon junction array predicts the quantum phase transition between superfluid and incoherent fluid states. Superfluid can be destroyed by a relatively weak repulsive interaction due to the lack of phase coherence, rather than the vanishing of order parameter.

\begin{figure}
\centering
\includegraphics[scale = 1.25]{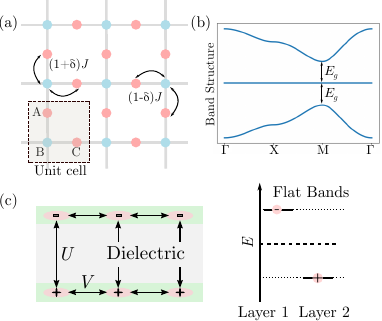}
\caption{(Adapted from Ref. \cite{ying2024flat}) (a) Lieb lattice hopping model: There are three orbitals in a unit cell. The intra-unit-cell hopping amplitude is $(1+\delta)J$, which is stronger than the inter-unit-cell hopping amplitude $(1-\delta)J$. (b) Band structure of Lieb lattice hopping model: There is an exactly flat band separated from two dispersive bands by an energy gap $E_g\propto \delta J$. Band structure is plotted along the momenta $\Gamma(0,0)\rightarrow \text{X}(\pi,0)\rightarrow \text{M}(\pi,\pi)\rightarrow\Gamma(0,0)$. (c) Lieb lattice bilayer with spatially indirect excitons. The attractive interaction between electron and hole is $U$, while the repulsive interaction between excitons is $V$. The electron and hole resides in the flat bands of two layers separately. When the interaction $U$ is large, there will be a finite density of excitons. The excitons may or may not condense to a superfluid phase, depending on the strength of the repulsive interaction.}
\label{Fig:LiebLattice_BandStructure_QuantumMetric}
\end{figure}

To demonstrate the points above, we take the Lieb lattice bilayer \cite{PhysRevLett.62.1201} as an example. A similar discussion on Lieb lattice and excitons can be found in Ref.~\cite{ying2024flat}. As shown in Fig.~\ref{Fig:LiebLattice_BandStructure_QuantumMetric}(a), Lieb lattice is defined on a square lattice, with three orbitals within a unit cell (as indicated by the gray box). The amplitudes for intra- and inter-unit-cell hoppings are $(1+\delta)J$ and $(1-\delta)J$. The dimensionless parameter $\delta$ is an important parameter for Lieb lattice model, as will be discussed below. The Bloch Hamiltonian is given by:
\begin{equation}
    H(\boldsymbol{k})=\begin{bmatrix}
	0 & a_{\boldsymbol{k}} & 0\\
	a_{\boldsymbol{k}}^* & 0 & b_{\boldsymbol{k}}\\
	0 & b_{\boldsymbol{k}}^* & 0
	\end{bmatrix}
    \label{Eq:LiebLattice_BlochH}
\end{equation}
with $a_{\boldsymbol{k}}=-J(1+\delta)-J(1-\delta)e^{ik_y}$ and $b_{\boldsymbol{k}}=-J(1+\delta)-J(1-\delta)e^{ik_x}$. Fig.~\ref{Fig:LiebLattice_BandStructure_QuantumMetric}(b) shows that this model supports an \emph{isolated, non-topological} flat band, separated from two dispersive bands by a finite energy gap $E_g\propto \delta J$. 

In this work, we focus on physics related to the flat bands. The periodic part of the Bloch wavefunction for the flat band is given by:
 \begin{equation}
    u_{\boldsymbol{k}}=\frac{1}{\mathcal{N}_{\boldsymbol{k}}}\left[-b_{\boldsymbol{k}},0,a_{\boldsymbol{k}}^*\right]^{\text{T}}
    \label{Eq:FB_BlochWF}
\end{equation}
with the normalization constant $\mathcal{N}_{\boldsymbol{k}}=\sqrt{\left|a_{\boldsymbol{k}}\right|^2+\left|b_{\boldsymbol{k}}\right|^2}$. One important feature of the flat band is that the flat band shows a widely tunable quantum metric \cite{provost1980riemannian,PhysRevB.56.12847,cheng2010quantum}. The tuning parameter is $\delta$. Quantum metric is defined as the real part of the quantum geometric tensor, $g_{\mu\nu}(\boldsymbol{k})=\text{Re} \left[\left(\partial_{\mu}u^{\dagger}_{\boldsymbol{k}}\right)\left(1-u_{\boldsymbol{k}}u^{\dagger}_{\boldsymbol{k}}\right)\left(\partial_{\nu}u_{\boldsymbol{k}}\right)\right]$, with $\partial_{\mu}=\partial_{k^{\mu}}$. One particular important quantity is the trace of Brillouin zone averaged quantum metric: $\text{tr}\ g_{\mu\nu}=\frac{(2\pi)^2}{\mathcal{A}_{\text{B.Z.}}}\int_{\text{B.Z.}}\frac{d^2k}{(2\pi)^2}\ \text{tr}\ g_{\mu\nu}(\boldsymbol{k})$, where $\mathcal{A}_{\text{B.Z.}}$ is the area of the first Brillouin zone. Direct calculation shows that the trace of averaged quantum metric diverges when $\delta$ approaches zero and vanishes when $\delta=1$. 


We consider a bilayer of Lieb lattice as shown in Fig.~\ref{Fig:LiebLattice_BandStructure_QuantumMetric}(c). The electrons and holes reside in the flat bands of the two layers separately. There is a strong attractive interaction $U$ that binds electron and hole into a boundstate of exciton. At the mean time, there is a much weaker repulsive interaction, $V$, between excitons. In the rest of the manuscript, we assume $V\ll U$ and ask whether the superfluidity of exciton condensate is stable under weak repulsive interaction.

To begin with, let us review the mean field result of exciton condensate. We assume that a mean field description is possible at \emph{zero} temperature in the \emph{absence} of repulsive interaction $V$. There are two facts to notice. One is that the mean field order parameter is on the order of attractive interaction. Second, superfluid stiffness is closely related to the quantum metric, namely properties of electronic wavefunctions.

For simplicity, we consider a contact attractive interaction $H_{\text{int}}=U\sum_{\boldsymbol{r}}\sum_{\alpha=\text{A,B,C}}c^{\dagger}_{\alpha}(\boldsymbol{r})d^{\dagger}_{\alpha}(\boldsymbol{r})d_{\alpha}(\boldsymbol{r})c_{\alpha}(\boldsymbol{r})$, where $c(d)^{\dagger}_{\alpha}(\boldsymbol{r})$ is the creation operator of an electron in layer 1(2) located at unit cell $\boldsymbol{r}$ and orbital $\alpha=\text{A,B,C}$; $c(d)_{\alpha}(\boldsymbol{r})$ is the corresponding annihilation operators. One can perform Hubbard-Stratonovich (HS) transformation to introduce the order parameter for exciton condensate:
\begin{equation}
    \begin{split}
    &H_{\text{int}}\rightarrow\\
    -&\sum_{\boldsymbol{r}}\sum_{\alpha} \left[\frac{1}{U}\left|\Delta_{\boldsymbol{r},\alpha}\right|^2 +\Delta_{\boldsymbol{r},\alpha}d^{\dagger}_{\alpha}(\boldsymbol{r})c_{\alpha}(\boldsymbol{r})+\bar{\Delta}_{\boldsymbol{r},\alpha}c^{\dagger}_{\alpha}(\boldsymbol{r})d_{\alpha}(\boldsymbol{r})\right]
    \end{split}
\end{equation}
The saddle point equation for the HS field $\Delta_{\boldsymbol{r},\alpha}$ and $\bar{\Delta}_{\boldsymbol{r},\alpha}$ defines the self-consistent mean field solution for the exciton condensate (superfluid) phase. 
The mean field analysis on exciton condensate closely mimics that of flat band superconductivity. Details can be found in Refs. \cite{PhysRevLett.132.026002,PhysRevB.101.060505,PhysRevB.106.014518} as well as SM \cite{SupplementalM}. 

\begin{figure}
\centering
\includegraphics[scale = 1.2]{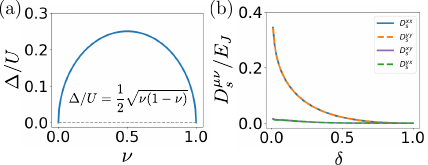}
\caption{(a) Mean field order parameter for exciton condensate. The mean field order parameter is indpendent of the parameter $\delta$ of Lieb lattice. It is on the order of the attractive interaction $U$ over a wide range of filling $\nu$. (b) Superfluid stiffness decreases monotonically with $\delta$. The SF stiffness is measured in units of $E_J=U\nu(1-\nu)$.}
\label{Fig:MF_OP_SS}
\end{figure}

The mean field order parameter turns out to be a function of electron filling. Here, we assume the filling of layer 1 is $\nu_1=\nu$ and layer 2 being $\nu_2=1-\nu$. Therefore, the total filling is $\nu_{\text{tot}}=1$ (only counting the filling of the flat bands). More explicitly, the mean field order parameter is given by:
\begin{equation}
    \Delta_{\alpha=\text{A,C}}=\Delta=\frac{U}{2}\sqrt{\nu(1-\nu)},\ \ \ \Delta_{\alpha=\text{B}}=0
    \label{Eq:MF_OrderParameter}
\end{equation}
As in Fig.~\ref{Fig:MF_OP_SS}(a), for a wide range of electron filling, the mean field order parameter is on the order of attractive interaction, $\Delta\sim U$. Meanwhile, $\Delta$ is maximal when each layer is at half filling. Mean field order parameter $\Delta$ vanishes when each layer approaches integer filling. 

Another feature to notice is that the mean field order parameter is independent of the model parameter $\delta$. Even for $\delta=1$, one can still find the same order parameter, Eq.~(\ref{Eq:MF_OrderParameter}). In this case, it is clear that there cannot be superfluidity, because unit cells are decoupled, Fig.~\ref{Fig:LiebLattice_BandStructure_QuantumMetric}(a). Therefore, a nonzero order parameter does not necessarily imply the global phase coherence. It turns out that one has to look at the dynamics of phase fluctuations. Below, we first review the properties of flat band superfluid stiffness. Then, we discretize the superfluid stiffness term on a lattice, which results in the model of Josephson junction array.

At the mean field level, the stability of exciton condensate is reflected in the finite superfluid stiffness $D_s^{\mu\nu}$. The SF order parameter has fluctuations in both the magnitude and phase $\Delta_{\boldsymbol{r}}\sim \left(\Delta_{MF}+\delta\Delta_{\boldsymbol{r}}\right)e^{i\varphi_{\boldsymbol{r}}}$. While the magnitude fluctuation is generally massive, the phase fluctuation $\varphi_{\boldsymbol{r}}$ is soft. The superfluid stiffness suppresses the spatial fluctuation in the phase $\varphi_{\boldsymbol{r}}$, making a globally phase coherent state possible. Superfluid stiffness can be obtained by considering response to an external gauge field or a response to a phase gradient in the mean field order parameter \cite{ying2024flat,PhysRevLett.132.026002,PhysRevB.101.060505,PhysRevLett.132.236001}. The superfluid stiffness shows up in a term in free energy that suppresses the spatial phase fluctuation:
\begin{equation}
    \begin{split}
        &\delta\mathcal{F}\sim D_s^{\mu\nu}\partial_{r_{\mu}}\varphi_{\boldsymbol{r}}\ \partial_{r_{\nu}}\varphi_{\boldsymbol{r}}\\
        &D_s^{\mu\nu}=U\nu(1-\nu)\int_{\text{BZ}}\frac{d^2k}{(2\pi)^2}g_{\mu\nu}(\boldsymbol{k}).
    \end{split}
    \label{Eq:SF_Stiffness}
\end{equation}
As shown in Fig.~\ref{Fig:MF_OP_SS}(b), the superfluid stiffness decreases with increasing value of $\delta$, suggesting that SF is more stable when quantum metric is large. The superfluid stiffness term as presented above is in the continuous description. On a lattice, such a term is replaced by a function of phase differences $\varphi_{\boldsymbol{r}}-\varphi_{\boldsymbol{r}^{\prime}}$ as discussed below.

\begin{figure}
\centering
\includegraphics[scale = 1]{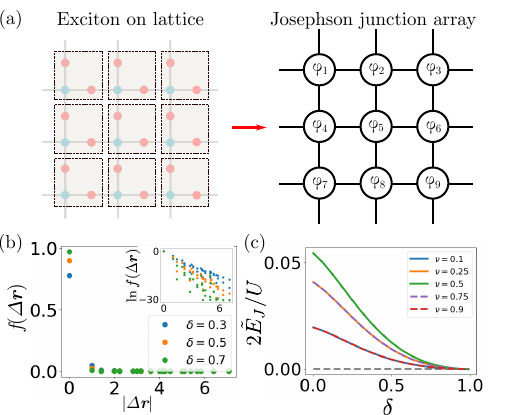}
\caption{(a) Mapping the exciton condensation on a lattice to an array of Josephson junction. Each unit cell is approximated as a superfluid island, characterized by a phase fluctuation $\varphi_{\boldsymbol{r}}$. Different unit cells are coupled by a Josephson coupling. (b) Strength of the Josephson coupling as a function of distance. The strength of JJ coupling, even though nonlocal, decreases exponentially with distance (inset). (c) Effective JJ coupling strength as a function of Lieb lattice parameter $\delta$ and electron filling $\nu$. The effective JJ coupling is large when the quantum metric is large (small $\delta$) and filling is close to one half ($\nu\sim\frac{1}{2}$).}
\label{Fig:Effective_Josephson}
\end{figure}

Discretizing the superfluid stiffness term on a lattice leads to an effective Josephson coupling between phase fluctuations $\varphi_{\boldsymbol{r}}$ at different unit cells, Fig.~\ref{Fig:Effective_Josephson}(a). Phenomenologically, one can write down the following Josephson coupling:
\begin{equation}
    \begin{split}
        &\hat{H}_J=-E_J\sum_{\boldsymbol{r}\neq\boldsymbol{r}^{\prime}}f(\boldsymbol{r},\boldsymbol{r}^{\prime})\cos\left(\varphi_{\boldsymbol{r}}-\varphi_{\boldsymbol{r}^{\prime}}\right)\\
        &f(\boldsymbol{r},\boldsymbol{r}^{\prime})=
        \int\frac{d^2k}{(2\pi)^2}\int\frac{d^2k^{\prime}}{(2\pi)^2}e^{i\left(\boldsymbol{k}-\boldsymbol{k}^{\prime}\right)\cdot\left(\boldsymbol{r}-\boldsymbol{r}^{\prime}\right)}\left|u^{\dagger}_{\boldsymbol{k}}u_{\boldsymbol{k}^{\prime}}\right|^2
    \end{split}
    \label{Eq:JJCoupling}
\end{equation}
This is the simplest symmetry-allowed term, given that the phase fluctuation $\varphi_{\boldsymbol{r}}$ is a compact boson. Without any competing fluctuation, the phase fluctuations at different positions tend to be synchronized, leading to a phase coherent superfluid phase. 

We should comment on the strength of the Josephson coupling $E_J$. One can identify $E_J=U\nu(1-\nu)$. In this way, a gradient expansion of the the phase difference $\varphi_{\boldsymbol{r}}-\varphi_{\boldsymbol{r}^{\prime}}\sim \left(\boldsymbol{r}-\boldsymbol{r}^{\prime}\right)\cdot\partial_{\boldsymbol{r}}\varphi_{\boldsymbol{r}}$ gives the superfluid stiffness. Notice that the function $f(\boldsymbol{r},\boldsymbol{r}^{\prime})$ depends on the periodic part of the Bloch wavefunction of the flat band $u_{\boldsymbol{k}}$. Its functional form gives the correct dependence of superfluid stiffness on the quantum metric.

The strength of the Josephson coupling highly depends on the model parameters. Fig.~\ref{Fig:Effective_Josephson}(b) plots the spatial profile of the function $f(\Delta\boldsymbol{r}=\boldsymbol{r}^{\prime}-\boldsymbol{r})=f(\boldsymbol{r},\boldsymbol{r}^{\prime})$ in Eq.~(\ref{Eq:JJCoupling}) describing the Josephson coupling. Fig.~\ref{Fig:Effective_Josephson}(b) shows that the Josephson coupling decays very fast as distance increases. Therefore, the energy scale for the phase dynamics is generally small, $E_Jf(\boldsymbol{r},\boldsymbol{r}^{\prime})|_{\boldsymbol{r}\neq\boldsymbol{r}^{\prime}}\ll E_J\sim U$. Given this scale separation, one should be able to focus on the dynamics associated with the phase $\varphi_{\boldsymbol{r}}$.

The inset of Fig.~\ref{Fig:Effective_Josephson}(b) shows that the Josephson coupling decays exponentially with distance. More importantly, the Josephson coupling is generally stronger with smaller value of $\delta$, namely larger quantum metric. Therefore, the superfluid phase is expected to be more stable in the regime with larger quantum metric.




The last piece of information for the problem is the density-density repulsive interaction. For demonstration purpose, we consider a simplest form of interaction:
\begin{equation}
    H_{C}=\frac{1}{2}V\sum_{\boldsymbol{r}}\left[\sum_{\alpha}c^{\dagger}_{\alpha}(\boldsymbol{r})c_{\alpha}(\boldsymbol{r})\right]^2+\frac{1}{2}V\sum_{\boldsymbol{r}}\left[\sum_{\alpha}d^{\dagger}_{\alpha}(\boldsymbol{r})d_{\alpha}(\boldsymbol{r})\right]^2
\end{equation}
This interaction follows from assuming each unit cell is similar to a quantum dot. There is local charging energy associated with each unit cell. Indeed, any density-density repulsive interaction can be thought of as an capacitive coupling. Here, we take the simplest situation. We would like to rewrite this repulsive interaction in the following form:
\begin{equation}
    H_{C}=\frac{1}{2}V\sum_{\boldsymbol{r}}\left[\sum_{\alpha}c^{\dagger}_{\alpha}(\boldsymbol{r})c_{\alpha}(\boldsymbol{r})-d^{\dagger}_{\alpha}(\boldsymbol{r})d_{\alpha}(\boldsymbol{r})\right]^2+\cdots
\end{equation}
The dots include interlayer interactions proportional to $V$, which are neglected due to the assumption $V\ll U$. The reason for such rewriting is that the operator
\begin{equation}
    \frac{1}{2}\Delta\hat{n}(\boldsymbol{r})=\frac{1}{2}\sum_{\alpha}\left[c^{\dagger}_{\alpha}(\boldsymbol{r})c_{\alpha}(\boldsymbol{r})-d^{\dagger}_{\alpha}(\boldsymbol{r})d_{\alpha}(\boldsymbol{r})\right]\sim -i\frac{\partial}{\partial\varphi_{\boldsymbol{r}}}
\end{equation}
is the canonical momentum of the phase fluctuation $\varphi_{\boldsymbol{r}}$. This fact is most easily seen by making a gauge transformation: $c_{\alpha}(\boldsymbol{r})\rightarrow e^{i\varphi_{\boldsymbol{r}}/2}c_{\alpha}(\boldsymbol{r})$ and $d_{\alpha}(\boldsymbol{r})\rightarrow e^{-i\varphi_{\boldsymbol{r}}/2}d_{\alpha}(\boldsymbol{r})$. The action, $\mathcal{S}$, now acquires a term $\sim \Dot{\varphi}_{\boldsymbol{r}}\Delta\hat{n}(\boldsymbol{r})/2$. Indeed, the operator $\Delta\hat{n}(\boldsymbol{r})$ couples to the time derivative of $\varphi_{\boldsymbol{r}}$. By definition, this is the canonical momentum of the phase fluctuation $\frac{1}{2}\Delta\hat{n}(\boldsymbol{r})=-i\frac{\partial}{\partial\varphi_{\boldsymbol{r}}}=\frac{\delta \mathcal{S}}{\delta\Dot{\varphi}_{\boldsymbol{r}}}$. More details can be found in SM \cite{SupplementalM}.

To this end, we are able to obtain a quantum Hamiltonian, that describes the competition between superfluidity and repulsive interaction:
\begin{equation}
    \hat{H}_{JJ}=-\frac{1}{2}E_C\sum_{\boldsymbol{r}}\frac{\partial^2}{\partial\varphi_{\boldsymbol{r}}^2}-E_J\sum_{\boldsymbol{r}\neq\boldsymbol{r}^{\prime}}f(\boldsymbol{r},\boldsymbol{r}^{\prime})\cos\left(\varphi_{\boldsymbol{r}}-\varphi_{\boldsymbol{r}^{\prime}}\right)
    \label{Eq:Hamiltonian_JJ_array}
\end{equation}
where $E_C=4V$. This Hamiltonian also describes the well-studied system of Josephson junction array \cite{tinkham2004introduction}. This model shows two phases. One is the phase coherent superfluid phase at large $E_J$. The other is the phase of incoherent fluid at large $E_C$. At large $E_J\gg E_C$, the phases $\varphi_{\boldsymbol{r}}$ is synchronized, because of the Josephson coupling. This is the phase of superfluid. In the opposite limit of large charging energy $E_C\gg E_J$, particle number $-i\partial_{\varphi_{\boldsymbol{r}}}$ is fixed, leading to a random/unsynchronized phase $\varphi_{\boldsymbol{r}}$. The resulting state is an incoherent fluid.

\begin{figure}
\centering
\includegraphics[scale = 1.2]{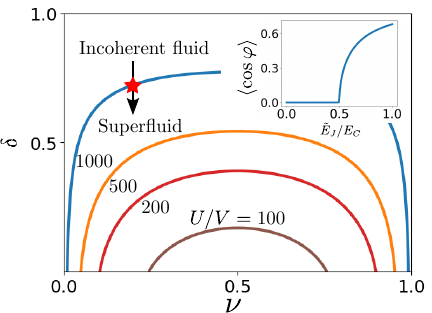}
\caption{Phase Diagram. Phase boundaries separating the phase of incoherent fluid and superfluid at various ratios between attractive and repulsive interactions $U/V=1000$ (blue), $500$ (orange), $200$ (red) and $100$ (brown). Inset: Phase diagram of Josephson junction array. Expectation value of $\langle\cos\varphi\rangle$ develops nonzero value when $\tilde{E}_J/E_C>\frac{1}{2}$.}
\label{Fig:CosExpectation}
\end{figure}

Indeed, a mean field type analysis shows that the critical point lies at $\tilde{E}_J=\frac{1}{2}E_C$ \cite{newrock2000two}. (The effective Josephon coupling $\tilde{E}_J$ is defined in the next paragraph.) The mean field Hamiltonian is given by:
\begin{equation}
\begin{split}
    &\hat{H}_{MF}=-\frac{1}{2}E_C\frac{\partial^2}{\partial\varphi^2}-\tilde{E}_J\langle\cos\varphi\rangle\cos\varphi\\
&\langle\cos\varphi\rangle=\int_{0}^{2\pi}d\varphi\cos\varphi\left|\Psi(\varphi)\right|^2
\end{split}
\end{equation}
where $\Psi(\varphi)$ is the wavefunction for the lowest energy level of the mean field Hamiltonian $\hat{H}_{MF}$. With proper phase choice, the expectation value $\langle\sin\varphi\rangle=0$ vanishes. The phase transition is reflected in the development of a nonzero expectation value of the cosine potential $\cos\varphi$ at the critical point $\tilde{E}_J=\frac{1}{2}E_C$ (Fig.~\ref{Fig:CosExpectation} Inset).

Notice that the effective Josephson coupling (summation is only over $\boldsymbol{r}^{\prime}$)
\begin{equation}
\Tilde{E}_J=\sum_{\boldsymbol{r}^{\prime}\neq\boldsymbol{r}}E_{J}f(\boldsymbol{r},\boldsymbol{r}^{\prime})
\end{equation}
is a function of model parameter, namely $\delta$ and electron filling $\nu$. Fig.~\ref{Fig:Effective_Josephson}(c)
shows that the effective Josephson coupling decreases with increasing $\delta$ at given electron filling. This is because increasing $\delta$ reduces the spatial quantum fluctuation (quantum metric). At the mean time, the effective Josephson coupling is maximum when both of the two layers of Lieb lattice are at half filling $\nu=\frac{1}{2}$. This is when the exciton density is maximum. Namely, the separation between excitons is minimum and a relatively strong coupling between excitons is expected.

Fig.~\ref{Fig:CosExpectation} plots the phase boundary between the incoherent fluid and the superfluid phase at various ratios of attractive interaction and the repulsive interaction $U/V$. Increasing the strength of repulsive interaction $V$ shrinks the regime of superfluidity. Take $U/V=1000$ as an example. Following the arrow, the system undergoes a quantum phase transition from an incoherent fluid to the phase of superfluid. Namely, the superfluid phase is established when the quantum metric is large (small $\delta$) and large exciton density $\nu\sim\frac{1}{2}$. In contrast, at small quantum metric $\delta\rightarrow1$, very weak repulsion is enough to destroy the superfluid behavior, even though a mean field order parameter $\Delta$ is nonzero. 

We should emphasize that the quantum phase transition along the arrow is totally driven by the change in the electronic wavefunction $u_{\boldsymbol{k}}$ of the flat band, Eq.~(\ref{Eq:FB_BlochWF}) and Eq.~(\ref{Eq:JJCoupling}) without changing the single electronic kinetic energy. Indeed, descreasing $\delta$ generally makes the Josephson coupling stronger, Fig.~\ref{Fig:Effective_Josephson}. Thus, superfluid phase is favored when $\delta$ is small. As mentioned, similar expectation can be drawn from the dependence of superfluid stiffness on the quantum metric Fig.~\ref{Fig:MF_OP_SS}(b). At the mean time, adding some dispersion favors the superfluid phase. Qualitatively, adding dispersion can be achieved by adding additional hopping terms to the model. The additional hopping terms essentially provide additional channel for exciton hopping and enhance the Josephson coupling, favoring the superfluid phase \cite{SupplementalM}.

The result presented above is really beyond mean field description. Indeed, as emphasized, over a wide range of filling, the mean field order parameter is on the order of attractive interaction $\Delta\sim U$. A weak repulsive interaction $V\ll U$ is unlikely to diminish the order parameter in a perturbative calculation. Nevertheless, such weak repulsive interaction is enough to poison the superfluid if quantum metric is small or exciton density is low.

Lastly, we point out that the result suggests that the superfluid stiffness can be renormalized strongly even with a weak repulsive interaction. Such renormaliztion cannot be captured by the coninuous description, Eq.~(\ref{Eq:SF_Stiffness}). Indeed, the continuous description of superfluid stiffness, Eq.~(\ref{Eq:SF_Stiffness}), together with quantum fluctuations, essentially describes a free boson. Namely, the action is given by $\mathcal{S}\sim \Dot{\varphi}\Pi-E_C\Pi^2-D_s(\partial_{\boldsymbol{r}}\varphi)^2$, where $\Pi$ is the canonical momentum. There is no interaction effect at this \emph{quadratic} level of description. Indeed, the generalization to Josephson coupling is necessary to capture the competition between repulsive interaction and the superfluidity. Namely, the superfluid stiffness diminishes upon approaching the phase boundary in Fig.~\ref{Fig:CosExpectation}, rendering superfluid to an incoherent fluid.


\emph{Acknowledgement}: We are grateful to H.C. Po, A. Kamenev, H. Goldman, Z. Song, D. Chowdhury, H. Wang, and M. Ye for discussions. X.Y. acknowledges the support of Hong Kong Research Grants Council (HKRGC) through Grant No. PDFS2425-6S02. K.L. is supported by HKRGC through C7037-22GF. 

\bibliography{ExcitonCondensate}

\end{document}


\title{Supplemental Material: Quantum-metric driven transition between superfluid and incoherent fluid}

\author{Xuzhe Ying} \email{yingxz@ust.hk}
\author{Kangle Li}

\affiliation{Department of Physics, Hong Kong University of Science and Technology, Clear Water Bay, Hong Kong, China}
\begin{abstract}
    We provide certain technical details to support the discussion in the main text. We first present the mean field analysis on the exciton condensate in the bilayer system. Then, we provide details on deriving the model of Josephson junction array from both phenomenological and microscopic perspectives. Lastly, we show that additional hopping terms (which gives the flat band a dispersion) enhances the Josephson coupling, namely favoring the superfluid phase.
\end{abstract}

\maketitle

\section{Mean field analysis}

The mean field study closely resembles and sometimes is a repetition of the analysis in Ref.~\cite{ying2024flat,PhysRevLett.132.026002,PhysRevLett.132.236001}. Nevertheless, let us provide some details here for completeness.

\subsection{Mean field Hamiltonian}

Let us start with the mean field Hamiltonian for the exciton condensation in the system of Lieb lattice bilayer:
\begin{equation}
    H_{\text{MF}}(\boldsymbol{k})=\begin{bmatrix}
        H(\boldsymbol{k})-\mu & \Delta_{\text{mat}}^{\dagger}\\
        \Delta_{\text{mat}} & H(\boldsymbol{k})+\mu
    \end{bmatrix}
\end{equation}
The mean field Hamiltonian above is written in the block form, where the blocks are labelled by layers. The diagonal blocks describes the free fermion model of the two layers respectively. The free fermion Hamiltonian $H(\boldsymbol{k})$ is given by Eq.~(1) in the maintext. The layer dependent potential $\pm\mu$ is chosen such that the total filling of the bilayer system is $\nu_1+\nu_2=1$ (only counting the filling the flat bands). The off-diagonal blocks is the mean field order parameter, which coherently couples the two layers. For a contact interaction such as Eq.~(3) in the maintext, the mean field order parameter is introduced as being diagonal in orbital basis $\Delta_{\text{mat}}=\text{Diag}\left[\Delta_{\text{A}},\Delta_{\text{B}},\Delta_{\text{C}}\right]$. Below, we will determine the order paremeter as well as the electron filling in a self-consistent manner.

It is useful to write the Green's function in the Bloch band basis of the Lieb lattice hopping model, namely $H(\boldsymbol{k})$. As discussed in the maintext, the Lieb lattice model contains one flat band and two dispersive bands. We denote the periodic part of Bloch wavefunctions as $u_{\text{f(d)},\boldsymbol{k}}$, where the sub-indices ``f(d)'' represents flat and dispersive bands respectively. We should mention that we write the Bloch wavefunctions in a very compact manner. Namely, $u_{\text{f},\boldsymbol{k}}$ should be understood as a $3\times 1$ matrix (corresponding to the \emph{one} flat band) and $u_{\text{d},\boldsymbol{k}}$ should be understood as a $3\times 2$ matrix (corresponding to the \emph{two} dispersive bands).

In the Bloch band basis of Lieb lattice hopping model, the explicit expression for the inverse Green's function for the bilayer system is given by:
\begin{equation}
    G^{-1}_{\text{MF}}(i\omega_m,\boldsymbol{k})=i\omega_m-\begin{bmatrix}
        H_{\text{f}}(\boldsymbol{k}) & \Delta_{\text{df}}^{\dagger}(\boldsymbol{k})\\
        \Delta_{\text{df}}(\boldsymbol{k}) & H_{\text{d}}(\boldsymbol{k})
    \end{bmatrix}
    \label{Eq:InvGreensFunction}
\end{equation}
where $i\omega_m$ is the Matsubara frequency. The inverse Green's function is separated into four blocks, corresponding to the subspaces of the flat bands and dispersive bands, as indicated by the sub-indices ``f(d)''. Each block is defined as:
\begin{equation}
    H_{\text{f}}(\boldsymbol{k})=\begin{bmatrix}
        -\mu & u^{\dagger}_{\text{f},\boldsymbol{k}}\Delta_{\text{mat}}^{\dagger}u_{\text{f},\boldsymbol{k}}\\
        u^{\dagger}_{\text{f},\boldsymbol{k}}\Delta_{\text{mat}}u_{\text{f},\boldsymbol{k}} & \mu
    \end{bmatrix}
\end{equation}
\begin{equation}
    H_{\text{d}}(\boldsymbol{k})=\begin{bmatrix}
        \epsilon_{\text{d}}(\boldsymbol{k})-\mu & u^{\dagger}_{\text{d},\boldsymbol{k}}\Delta_{\text{mat}}^{\dagger}u_{\text{d},\boldsymbol{k}}\\
        u^{\dagger}_{\text{d},\boldsymbol{k}}\Delta_{\text{mat}}u_{\text{d},\boldsymbol{k}} & \epsilon_{\text{d}}(\boldsymbol{k})+\mu
    \end{bmatrix}
\end{equation}
\begin{equation}
    \Delta_{\text{df}}(\boldsymbol{k})=\begin{bmatrix}
        0 & u^{\dagger}_{\text{d},\boldsymbol{k}}\Delta_{\text{mat}}^{\dagger}u_{\text{f},\boldsymbol{k}}\\
        u^{\dagger}_{\text{d},\boldsymbol{k}}\Delta_{\text{mat}}u_{\text{f},\boldsymbol{k}} & 0
    \end{bmatrix}
\end{equation}
For clarification, we should mention that $H_{\text{f}}(\boldsymbol{k})$, $H_{\text{d}}(\boldsymbol{k})$, and $\Delta_{\text{df}}(\boldsymbol{k})$ are $2\times2$, $4\times4$, and $4\times2$ matrices respectively.

\subsection{Mean field equations: order parameter and filling}

Given the mean field Hamiltonian as well as the inverse Green's function, we are ready to write down the mean field equation. Indeed, the free energy density is formally given by:
\begin{equation}
    \mathcal{F}=\sum_{\alpha=\text{A,B,C}}\frac{\left|\Delta_{\alpha}\right|^2}{U}-\frac{1}{V\beta}\text{Tr}\ln G_{\text{MF}}^{-1}(i\omega_m,\boldsymbol{k})
\end{equation}
where $\beta$ and $V$ is the inverse of temperature and the system size respectively. The inverse Green's function $G_{\text{MF}}^{-1}(i\omega_m,\boldsymbol{k})$ is given by Eq.~(\ref{Eq:InvGreensFunction}). The free energy density formally follows from the partition function:
\begin{equation}
    \mathcal{Z}=\int D\bar{\psi} D\psi e^{-\frac{1}{U}\int_{0}^{\beta}d\tau\int d^2r\sum_{\alpha}\left|\Delta_{\alpha}\right|^2+\int_0^{\beta}d\tau\int d^2r\bar{\psi}G_{\text{MF}}^{-1}\psi}
\end{equation}
The free energy density is given by $\mathcal{F}=-\frac{1}{V\beta}\ln \mathcal{Z}$.

The mean field equations can be obtained by varying the free energy density:
\begin{equation}
    \frac{\delta \mathcal{F}}{\delta \bar{\Delta}_{\alpha}}=0,\ \ \frac{\delta \mathcal{F}}{\delta \mu}=n_1-n_2
\end{equation}
where the second equation determines the electron density difference of the two Lieb lattice layers, $n_1-n_2$.

Throughout this work, we assume a large energy gap between the dispersive bands and the flat band, $E_g\gg U$ (see Fig.~1(b) in maintext). For the purpose of this subsection, we can neglect the dispersive bands. In the next subsection, the dispersive bands are not negligible when we consider the superfluid stiffness.

We restrict ourselves to the flat bands only. The explicit expressions for the mean field equation reads:
\begin{equation}
    \begin{split}
        &\frac{\Delta_{\alpha}}{U}=\frac{1}{\beta}\sum_{i\omega_m}\int\frac{d^2k}{(2\pi)^2}\text{tr}\ G_{\text{f}}(i\omega_m,\boldsymbol{k})\begin{bmatrix}
            0 & -\left|u^{\alpha}_{\text{f},\boldsymbol{k}}\right|^2\\
            0 & 0
        \end{bmatrix}\\
        &\nu_1-\nu_2=\frac{1}{\beta}\sum_{i\omega_m}\int\frac{d^2k}{(2\pi)^2}\text{tr}\ G_{\text{f}}(i\omega_m,\boldsymbol{k})\begin{bmatrix}
            -1 & 0\\
            0 & 1
        \end{bmatrix}
    \end{split}
\end{equation}
There are several details to notice. First, the Green's function is defined as $G_{\text{f}}(i\omega_m, \boldsymbol{k})=\left(i\omega_m-H_{\text{f}}(\boldsymbol{k})\right)^{-1}$. Second, in the first equation, the factor $u^{\alpha}_{\text{f},\boldsymbol{k}}$ denotes the $\alpha$-th component of the Bloch wavefunction $u_{\text{f},\boldsymbol{k}}$. Lastly, $\nu_{1,2}$ are the electron filling of the flat bands only of the two layers separately. One can compute the flat band filling of the two layers separately and find that $\nu_1+\nu_2=1$. Below, we set:
\begin{equation}
    \nu_1=\nu,\ \ \ \nu_2=1-\nu
\end{equation}
Hence, $\nu_1-\nu_2=2\nu-1$.

Further simplification on the mean field equations can be made by performing the Matsubara summation and by assuming zero temperature. After Matsubara summation, the mean field equations at zero temperature reads:
\begin{equation}
    \begin{split}
        &\frac{\Delta_{\alpha}}{U}=\int\frac{d^2k}{(2\pi)^2}\frac{u^{\dagger}_{\text{f},\boldsymbol{k}}\Delta_{\text{mat}}u_{\text{f},\boldsymbol{k}}}{2\sqrt{\mu^2+\left|u^{\dagger}_{\text{f},\boldsymbol{k}}\Delta_{\text{mat}}u_{\text{f},\boldsymbol{k}}\right|^2}}\left|u^{\alpha}_{\text{f},\boldsymbol{k}}\right|^2\\
        &2\nu-1=-\int\frac{d^2k}{(2\pi)^2}\frac{\mu}{\sqrt{\mu^2+\left|u^{\dagger}_{\text{f},\boldsymbol{k}}\Delta_{\text{mat}}u_{\text{f},\boldsymbol{k}}\right|^2}}
    \end{split}
\end{equation}
One find the self-consistent solution in the system of Lieb lattice bilayer as:
\begin{equation}
\begin{split}
        &\Delta_{\text{A}} = \Delta_{\text{C}} = \frac{U}{2}\sqrt{\nu(1-\nu)}, \ \ \ \Delta_{\text{B}} = 0 \\
        &\mu = \frac{U}{4}(2\nu-1).
\end{split}
\label{Eq:MFSolution}
\end{equation}
As emphasized in the main text, the mean field order parameter is independent of the model parameter $\delta$. Indeed, the order parameter only depends on the interaction strength $U$ and the filling $\nu$.

\subsection{Superfluid stiffness}


To study the superfluid stiffness, we consider the situation when the order parameter acquires extra phase gradient $\Delta_{\boldsymbol{r},\alpha}=\Delta_\alpha e^{i\vec{Q}\cdot\vec{r}}$. The phase gradient $\boldsymbol{Q}$ leads to a dissipationless counterflow current \cite{ying2024flat}. To begin with, we consider a slightly different mean field Hamiltonian:
\begin{equation}
    H_{\rm MF}(\boldsymbol{k}-\boldsymbol{A})=\left[
    \begin{array}{cc}
       H(\boldsymbol{k}-\boldsymbol{A}_1)-\mu  & \Delta_{\text{mat}}^{\dagger} \\
       \Delta_{\text{mat}}  & H(\boldsymbol{k} -\boldsymbol{Q} - \boldsymbol{A}_2)+\mu \\
    \end{array}\right],
\end{equation}
where two auxiliary gauge fields $\boldsymbol{A}_{1,2}$ are introduced to aid finding the currents in both layers. 

In terms of partition function $\mathcal{Z}[\boldsymbol{A}]$ with background gauge field $\boldsymbol{A}_{1,2}$, 
\begin{equation}
    \mathcal{Z}[\boldsymbol{A}] = \int D\bar\psi D\psi e^{\int_0^\beta d\tau \int \frac{d^2k}{(2\pi)^2}\bar\psi(\boldsymbol{k},\tau)(-\partial_\tau - H_{MF}(\boldsymbol{k} -\boldsymbol{A}))\psi(\boldsymbol{k},\tau) }
\end{equation}
the free energy is 
\begin{equation}
    \mathcal{F}[\boldsymbol{A}] = -\frac{1}{\beta}\ln \mathcal{Z}[\boldsymbol{A}].
\end{equation}
The counterflow current $\boldsymbol{j}^{\CF}_l$ can be computed by taking the derivative with respect to the gauge field $\boldsymbol{A}_l$, 
\begin{equation}
    \boldsymbol{j}^{\CF}_l = -\beta\frac{\delta \mathcal{F}[\boldsymbol{A}]}{\delta A_l}{\Bigg |}_{\boldsymbol{A}_1=\boldsymbol{A}_2=0}
\end{equation}
To the linear order at $\boldsymbol{Q}$, the counterflow current is closely related to the quantum metric $\boldsymbol{j}^{\CF}_{l,\mu} \propto g_{\mu\nu}Q^\nu$.
In terms of Green's function $G_{\text{MF}}(i\omega_m,\boldsymbol{k}) = (i\omega_m-H_{\text{MF}}(\boldsymbol{k}))^{-1}$, the CF current is 
\begin{equation}\label{eq:current_CF}
    \boldsymbol{j}_l^{\CF} = -\frac{1}{\beta}\sum_{i\omega_m}\int \frac{d^2k}{(2\pi)^2}\text{tr} \left[G_{\text{MF}}(i\omega_m,\boldsymbol{k}) \hat{\boldsymbol{j}}_l(\boldsymbol{k})\right],
\end{equation}
with
\begin{equation}
    \hat{\boldsymbol{j}}_{1} = 
    -\left[\begin{array}{cc}
        \partial_{\boldsymbol{k}}H(\boldsymbol{k}) & 0 \\
        0 & 0 
    \end{array}\right],
    \hat{\boldsymbol{j}}_{2} = 
    -\left[\begin{array}{cc}
        0 & 0 \\
        0 & \partial_{\boldsymbol{k}}H(\boldsymbol{k}-\boldsymbol{Q})\\ 
    \end{array}\right].
\end{equation}
Instead of projecting to the flat band subspace, we need to consider all bands here. To compute $\boldsymbol{j}_{1/2}$, it is convenient to work in the Bloch band basis as in the previous subsections. We denote the flat band by ``f'' and dispersive bands by ``d''. The inverse Green's function under this basis is written as
\begin{equation}
    G_{\text{MF}}^{-1}(i\omega_m,\boldsymbol{k}) =i\omega_m - 
    \begin{bmatrix}
        H_{\rm f}(\boldsymbol{k}) & \Delta_{\rm df}^{\dagger}(\boldsymbol{k}) \\
        \Delta_{\rm df}(\boldsymbol{k}) & H_{\rm d}(\boldsymbol{k})
    \end{bmatrix},
\end{equation}
where the mean field Hamiltonian is written in the block form:
\begin{equation}
    \begin{split}
    H_{\rm f}(\boldsymbol{k}) &= \left[
    \begin{array}{cc}
        -\mu & u^\dagger_{{\rm f},\boldsymbol{k}}\Delta_{\text{mat}}^{\dagger} u_{{\rm f},\boldsymbol{k}-\boldsymbol{Q}} \\
        u^\dagger_{{\rm f},\boldsymbol{k}-\boldsymbol{Q}}\Delta_{\text{mat}} u_{{\rm f},\boldsymbol{k}} & \mu
    \end{array}
    \right],
     \\
    H_{\rm d}(\boldsymbol{k}) &= \left[
    \begin{array}{cc}
        \epsilon_{\rm d}(\boldsymbol{k})-\mu & u^\dagger_{{\rm d},\boldsymbol{k}}\Delta_{\text{mat}}^{\dagger} u_{{\rm d},\boldsymbol{k}-\boldsymbol{Q}} \\
        u^\dagger_{{\rm f},\boldsymbol{k}-\boldsymbol{Q}}\Delta_{\text{mat}} u_{{\rm f},\boldsymbol{k}} & \epsilon_{\rm d}(\boldsymbol{k})+\mu \\
    \end{array}
    \right],
    \\
    \Delta_{\rm df}(\boldsymbol{k}) &=
    \begin{bmatrix}
        0 & u_{{\rm d},\boldsymbol{k}}^\dagger\Delta^{\dagger}_{\text{mat}} u_{{\rm f},\boldsymbol{k}-\boldsymbol{Q}} \\
        u_{{\rm d},\boldsymbol{k}-\boldsymbol{Q}}^\dagger\Delta_{\text{mat}} u_{{\rm f},\boldsymbol{k}} & 0
    \end{bmatrix}.
    \end{split}
\end{equation}

To proceed, we assume a large gap between dispersive bands and the flat bands $|\epsilon_{\text{d}}|\gg|\Delta_m|$. To the order of $\mathcal{O}(\frac{1}{\epsilon_{\text{d}}})$, the Green's function $G_{\text{MF}}(i\omega_m,\boldsymbol{k})$ is given by (the dependence on $(i\omega_m,\boldsymbol{k})$ is assumed and is not written explicitly):
\begin{equation}
    G_{\text{MF}} = \left[
    \begin{array}{cc}
       G_{\rm f}+G_{\rm f}\Delta_{\rm df}^{\dagger}G_{\rm d}\Delta_{\rm df} G_{\rm f}  & -G_{\rm f}\Delta_{\rm df}^{\dagger}G_{\rm d} \\
       -G_{\rm d}\Delta_{\rm df}G_{\rm f}  & G_{\rm d}
    \end{array}
    \right]
\end{equation}
with $G_{\rm f/d} = (i\omega_m-H_{\rm f/d})^{-1}$. 

Next, we should write the current operator in the same Bloch band basis:
\begin{equation}
    \hat{\boldsymbol{j}}_{1}= \left[
    \begin{array}{cc}
        0 & \boldsymbol{j}_{1,{\rm fd}} \\
        \boldsymbol{j}^\dagger_{1,{\rm fd}} & \boldsymbol{j}_{\rm d}
    \end{array}
    \right]
\end{equation}
with 
\begin{equation}
\boldsymbol{j}_{1,\fd} = \left[
\begin{array}{cc}
    -u^\dagger_{\f,\boldsymbol{k}}\partial_{\boldsymbol{k}}H(\boldsymbol{k})u_{\rmd,\boldsymbol{k}} & 0 \\
    0 & 0
\end{array}
\right],\ 
\boldsymbol{j}_{1,\rmd} = -{\rm Diag}[\partial_{\boldsymbol{k}}\epsilon_\rmd(\boldsymbol{k})].
\end{equation}
where the diagonal elements are (the negative of) the group velocities of each band. Meanwhile, the off-diagonal elements is on the order of the gap $\epsilon_{\text{d}}$. Indeed, the off-diagonal terms can be written as
\begin{equation}
\boldsymbol{j}_{1,\rm fd} = \left[
\begin{array}{cc}
        (\partial_{\boldsymbol{k}}u_{\f,\boldsymbol{k}}^\dagger) u_\rmd \epsilon_d  &  0 \\
    0 & 0
\end{array}\right].
\end{equation}
The expression of the CF current can be derived by putting expressions of $G_0$ and $\hat{j}_{1}$ into Eq.\ \ref{eq:current_CF},
\begin{equation}
    \boldsymbol{j}_1^{\CF} =  \frac{1}{\beta}\sum_{i\omega_m}\int \frac{d^2k}{(2\pi)^2}\text{tr} \left[G_\f \Delta_{\text{df}}^{\dagger} G_\rmd \hat{\boldsymbol{j}}_{1,\fd}^\dagger + {\rm h.c.} \right],
\end{equation}
where the contribution from dispersive bands ($\sim\text{tr}\ G_{\text{d}}\boldsymbol{j}_{\text{d}}$) is ignored. This because at zero temperature the dispersive bands are either fully occupied or fully empty and thus have vanishing contribution to the current. 

Finally, the expression for the counterflow current reduces to
\begin{equation}
\begin{split}
    \boldsymbol{j}_1^{\CF} &= \int\frac{d^2k}{(2\pi)^2} \frac{1}{2\sqrt{\mu^2+\left|u^{\dagger}_{\text{f},\boldsymbol{k}}\Delta_{\text{mat}}u_{\text{f},\boldsymbol{k}}\right|^2}} \\
    & \times u^\dagger_{\f,\boldsymbol{k}}\Delta_m u_{\f,\boldsymbol{k}-\boldsymbol{Q}}u_{\f,\boldsymbol{k}-\boldsymbol{Q}}^\dagger \Delta_m^\dagger(\id-u_{\f,\boldsymbol{k}}u_{\f, \boldsymbol{k}}^\dagger)(\partial_{\boldsymbol{k}} u_{\f,\boldsymbol{k}}) + {\rm h.c.}
\end{split}
\end{equation}
If we employ the mean field solution Eq.~(\ref{Eq:MFSolution}), and take the small phase gradient limit $\left|\boldsymbol{Q}\right|\ll1$, the counterflow current is given by 
\begin{equation}
    j_{1,\mu}^{\CF} = U\nu(1-\nu)\int_{\rm BZ} \frac{d^2k}{(2\pi)^2}g_{\mu\nu}(\boldsymbol{k}) Q^\nu.
\end{equation}
where the quantum metric tensor is defined as $g_{\mu\nu}(\boldsymbol{k})=\text{Re} \left[\left(\partial_{\mu}u^{\dagger}_{\text{f},\boldsymbol{k}}\right)\left(1-u_{\text{f},\boldsymbol{k}}u^{\dagger}_{\text{f},\boldsymbol{k}}\right)\left(\partial_{\nu}u_{\text{f},\boldsymbol{k}}\right)\right]$, with $\partial_{\mu}=\partial_{k^{\mu}}$. Indeed, a phase gradient can drive a dissipationless current.

The result of the dissipationless counterflow current suggests that the phase gradient $\boldsymbol{Q}$ introduces a correction to the free energy as:
\begin{equation}
    \delta\mathcal{F}=\frac{1}{2}U\nu(1-\nu)\int_{\rm BZ} \frac{d^2k}{(2\pi)^2}g_{\mu\nu}(\boldsymbol{k}) Q^{\mu}Q^\nu
    \label{Eq:FE_SS}
\end{equation}
One can extract the superfluid stiffness as being proportional to the quantum metric:
\begin{equation}
    D_s^{\mu\nu}=U\nu(1-\nu)\int_{\rm BZ} \frac{d^2k}{(2\pi)^2}g_{\mu\nu}(\boldsymbol{k}).
\end{equation}

\subsection{Effects of band dispersion}

In reality, the electronic bands always have some dispersion. In this subsection, we study the correction to the superfluid stiffness from the dispersion of nearly flat bands. As shown in Ref.~\cite{PhysRevLett.132.236001}, there is additional diamagnetic contribution from the band dispersion.

To study the effects the band dispersion, we consider the following mean field Hamiltonian:
\begin{equation}\label{eq:flat_dispersive}
    \begin{split}
        H_{\rm MF}(\boldsymbol{k}-\boldsymbol{A})=&
    \begin{bmatrix}
        H(\boldsymbol{k}-\boldsymbol{A}_1)-\mu  & \Delta_{\text{mat}}^{\dagger} \\
       \Delta_{\text{mat}}  & H(\boldsymbol{k} -\boldsymbol{Q} - \boldsymbol{A}_2)+\mu
    \end{bmatrix}\\
    +&\begin{bmatrix}
        \epsilon(\boldsymbol{k}-\boldsymbol{A}_1)\mathbb{I} & 0\\
        0 & -\epsilon(\boldsymbol{k}-\boldsymbol{Q}-\boldsymbol{A}_2)\mathbb{I}
    \end{bmatrix}
    \end{split}
\end{equation}
Namely, we add identity matrices to the diagonal blocks of the mean field Hamitonian. The coefficients, $\pm\epsilon(\boldsymbol{k})$, makes the flat band dispersive. Such dispersion can be obtained from adding hopping between AA (and BB, CC at the same time) sites of adjacent unit cells.

The computation of the superfluid stiffness fully parallels that of the previous subsection. The major difference is that the current operator is now given by:
\begin{equation}
\hat{\boldsymbol{j}}_{1}= 
    \begin{bmatrix}
        \boldsymbol{j}_{\f} & \boldsymbol{j}_{1,{\rm fd}} \\
        \boldsymbol{j}^\dagger_{1,{\rm fd}} & \boldsymbol{j}_{\rm d}
    \end{bmatrix}
    , \boldsymbol{j}_{\f}=\begin{bmatrix}
        \partial_{\boldsymbol{k}}\epsilon(\boldsymbol{k}) & 0\\
        0 & 0
    \end{bmatrix}.
\end{equation}
Correspondingly, the counterflow current has a ``conventional'' contribution from the finite group velocity of the nearly flat bands:
\begin{equation}
    \boldsymbol{j}_1^{\text{CF,conv}}=\frac{1}{\beta}\sum_{i\omega_m}\int\frac{d^2k}{(2\pi)^2}\text{tr}\ G_{\text{f}}(i\omega_m,\boldsymbol{k})\boldsymbol{j}_{\text{f}}(\boldsymbol{k})
\end{equation}
in addition to the geometric contribution derived from previous subsection. We should mention that the geometric contribution to the CF current is more complicated as derived in Ref.~\cite{PhysRevLett.132.236001}. Simplifications can be made in our situation. Namely, the electronic states have the same Bloch wavefunction for both layers. Meanwhile, the mean field order parameter takes a relatively simple form, Eq.~(\ref{Eq:MFSolution}).

Direct calculation shows that the conventional contribution of the CF current is the diamagnetic current. The explicit expression is given by:
\begin{equation}
    j^{\text{CF,conv}}_{1,\mu} = -\int \frac{d^2k}{(2\pi)^2} \frac{\epsilon(\boldsymbol{k})-\mu}{4E(\boldsymbol{k})}\frac{\partial^2 \epsilon(\boldsymbol{k})}{\partial k_{\mu}\partial k_{\nu}} Q^{\nu}.
\end{equation}
where $E(\boldsymbol{k})=\sqrt{\left[\epsilon(\boldsymbol{k})-\mu\right]^2+\left|u^{\dagger}_{\text{f},\boldsymbol{k}}\Delta_{\text{mat}}u_{\text{f},\boldsymbol{k}}\right|^2}$.

Following the result above, the superfluid stiffness acquires a conventional/diamagnetic contribution from the band dispersion:
\begin{equation}
    D_{s,\text{conv}}^{\mu\nu}=-\int \frac{d^2k}{(2\pi)^2} \frac{\epsilon(\boldsymbol{k})-\mu}{4E(\boldsymbol{k})}\frac{\partial^2 \epsilon(\boldsymbol{k})}{\partial k_{\mu}\partial k_{\nu}}
\end{equation}
This term is always positive. To see this point, it's easiest to perform an integration by parts: $D_{s,\text{conv}}^{\mu\nu}=\int \frac{d^2k}{(2\pi)^2} \frac{\left|\Delta\right|^2}{4E^3(\boldsymbol{k})}\partial_{k_{\mu}} \epsilon(\boldsymbol{k})\partial_{k_{\nu}} \epsilon(\boldsymbol{k})$. Then, notice that the superfluid stiffness has the following matrix structure: $D^{\mu\nu}\sim\sum_i\begin{bmatrix}
        x_i^2 & x_iy_i\\
        x_iy_i & y_i^2
    \end{bmatrix}$,
which is a semi-positive-definite matrix. 

Thus, superfluid stiffness increases upon adding dispersion to the flat band. Notice that we added the dispersion in a very special way, such that the zero momentum excitons condense first. Namely, the condensate carries zero momentum. Qualitatively, adding dispersion to the flat bands can help stabilize superfluid phase.

\section{Model of Josephson junction array}
In this section we present the details of obtaining the effective Hamiltonian of Josephson junction array \cite{newrock2000two,tinkham2004introduction}. We first discuss the discretization of the superfluid term on a lattice, by comparing the Josephson coupling and superfluid stiffness term. Then, we discuss the quantum fluctuations introduced by the repulsive interaction. Lastly, we obtain the quantum Hamiltonian of Josephson junction array, which allows us to study the competition between (weak) repulsive interaction and the superfluidity.

\subsection{Josephson coupling}
As described in the main text, the phenomenological coupling is in the form
\begin{equation}\label{eq:jj_term}
\begin{split}
    \hat H_J &= -E_J\sum_{\boldsymbol{r}\neq\boldsymbol{r}'} f(\boldsymbol{r},\boldsymbol{r}')\cos(\varphi_{\boldsymbol{r}}-\varphi_{\boldsymbol{r}'}) \\
    f(\boldsymbol{r},\boldsymbol{r}') &= \int\frac{d^2k}{(2\pi)^2}\int\frac{d^2k'}{(2\pi)^2}e^{i(\boldsymbol{k}-\boldsymbol{k}')\cdot (\boldsymbol{r}-\boldsymbol{r}')} \left|u_{\boldsymbol{k}}^\dagger u_{\boldsymbol{k}'}\right|^2.
\end{split}
\end{equation}
The spatial profile of the Josephson coupling, as dictated by the function $f(\boldsymbol{r},\boldsymbol{r}^{\prime})$, is motivated by a more microscopic study of the correlation function (as detailed in the next section). 

Now, we need to determine the coefficient $E_J$. Expand $\hat H_J$ in powers of $\boldsymbol{r}-\boldsymbol{r}'$ and keep to the quadratic terms,
\begin{equation}
\begin{split}
    H_{(2)} &\approx -E_J\sum_{\boldsymbol{r}\neq\boldsymbol{r}'} f(\boldsymbol{r},\boldsymbol{r}') \cos\left( (\boldsymbol{r}-\boldsymbol{r}')\cdot\nabla_{\boldsymbol{r}}\varphi_{\boldsymbol{r}} \right) \\
    &= \frac{1}{2} E_J \sum_{\boldsymbol{r}\neq\boldsymbol{r}'} f(\boldsymbol{r},\boldsymbol{r}')(\boldsymbol{r}-\boldsymbol{r}')_\mu(\boldsymbol{r}-\boldsymbol{r}')_\nu \partial_{r_{\mu}}\varphi_{\boldsymbol{r}}\partial_{r_{\nu}} \varphi_{\boldsymbol{r}}.
\end{split}
\end{equation}
where constant term is neglected. The difference $(\boldsymbol{r}-\boldsymbol{r}')_\mu$ can be translated to the derivative of momentum $\boldsymbol{k}^{(\prime)}_\mu$ as follows,
\begin{equation}
\begin{split}
    &f(\boldsymbol{r},\boldsymbol{r}')(\boldsymbol{r}-\boldsymbol{r}')_\mu(\boldsymbol{r}-\boldsymbol{r}')_\nu \\
    =& \int \frac{d^2k}{(2\pi)^2}\frac{d^2k'}{(2\pi)^2} \left[\partial_{k_\mu}\partial_{k^{\prime}_\nu}e^{i(\boldsymbol{k}-\boldsymbol{k}')\cdot (\boldsymbol{r}-\boldsymbol{r}')} \right]\left|u^\dagger_{\boldsymbol{k}}u_{\boldsymbol{k}'}\right|^2 \\
    =& \int \frac{d^2k}{(2\pi)^2}\frac{d^2k'}{(2\pi)^2} e^{i(\boldsymbol{k}-\boldsymbol{k}')\cdot (\boldsymbol{r}-\boldsymbol{r}')} \partial_{k_\mu}\partial_{k^{\prime}_\nu} \left|u^\dagger_{\boldsymbol{k}}u_{\boldsymbol{k}'}\right|^2 \\
\end{split}
\end{equation}
The last line follows from integration by parts. After explicitly performing the momentum derivative on the Bloch wavefunctions, we are able to obtain the following Hamiltonian:
\begin{equation}
    \hat H_J = E_J\sum_{\boldsymbol{r}}\int\frac{d^2k}{(2\pi)^2}g_{\mu\nu}(\boldsymbol{k})\partial_{r_{\mu}}\varphi_{\boldsymbol{r}}\partial_{r_{\nu}}\varphi_{\boldsymbol{r}}
\end{equation}
By identifying $\partial_{r_{\mu}}\varphi_{\boldsymbol{r}}=Q^{\mu}$, one should recognize this term as the analog of the superfluid stiffness in the continuous description. Comparing with Eq.~(\ref{Eq:FE_SS}), we can identify
\begin{equation}
    E_J = 2U\nu(1-\nu).
    \label{Eq:EJ_Pheno}
\end{equation}

Before proceeding, we should comment on the correction from adding some dispersion to the flat bands. As discussed in the previous section, adding some dispersion to the flat band enhances the superfluid stiffness. In turn, one would expect the Josephson coupling also receives some correction, which further enhances the phase coherence. In Secion.~\ref{Sec:JJCoupling_AddHopping}, we show this is indeed the case.

\subsection{Repulsive interaction and quantum fluctuations}
Next we consider the quantum fluctuations introduced by the repulsive interaction. We consider the simplest repulsive interaction:
\begin{equation}
    H_c = \frac{V}{2}\sum_{\boldsymbol{r}}\left[\sum_{\alpha} c^\dagger_{\alpha}(\boldsymbol{r})c_\alpha(\boldsymbol{r})-d_{\alpha}^\dagger(\boldsymbol{r})d_{\alpha}(\boldsymbol{r})\right]^2,
\end{equation}
As emphasize in the maintext, the operator in the bracket is the canonical momentum of the phase fluctuation $\varphi_{\boldsymbol{r}}$:
\begin{equation}
    \frac{1}{2}\Delta\hat{n}(\boldsymbol{r})=\frac{1}{2}\sum_{\alpha}\left[c^{\dagger}_{\alpha}(\boldsymbol{r})c_{\alpha}(\boldsymbol{r})-d^{\dagger}_{\alpha}(\boldsymbol{r})d_{\alpha}(\boldsymbol{r})\right]\sim -i\frac{\partial}{\partial\varphi_{\boldsymbol{r}}}
\end{equation}
We should mention that we are using the fermionic operators $c(d)^{(\dagger)}$ (as in the maintext) and $\psi_{1(2)}$ ($\bar{\psi}_{1(2)}$) (as in the partition functions) interchangeably.

Assume the following a phase fluctuation in the order parameter $\Delta_{\boldsymbol{r}\alpha}e^{i\varphi_{\boldsymbol{r}}(t)}$, where the time dependence is written explicitly. As a reference for the analysis below, let us write down the Lagrangian in \emph{real} time path integral fomulation: ($\sigma_z$ is the Pauli matrix in layer basis)
\begin{equation}
    \begin{split}
        \mathcal{L}=&\bar{\psi}\left[i\partial_t-H_0\right]\psi -\frac{V}{2}\left(\sum_{\alpha}\bar{\psi}_{\alpha}\sigma_z\psi_{\alpha}\right)^2\\
        +&\sum_{\alpha}\Delta_{\boldsymbol{r},\alpha}e^{i\varphi_{\boldsymbol{r}}(t)}\bar{\psi}_{2,\alpha}(\boldsymbol{r})\psi_{1,\alpha}(\boldsymbol{r})+\bar{\Delta}_{\boldsymbol{r},\alpha}e^{-i\varphi_{\boldsymbol{r}}(t)}\bar{\psi}_{1,\alpha}(\boldsymbol{r})\psi_{2,\alpha}(\boldsymbol{r})
    \end{split}
\end{equation}
Next step is to perform a gauge transformation: 
\begin{equation}
\begin{split}
    \psi_{\boldsymbol{r}\alpha,1} &\to e^{i\varphi_{\boldsymbol{r}}(t)/2}\psi_{\boldsymbol{r}\alpha,1} \\
    \psi_{\boldsymbol{r}\alpha,2} &\to e^{-i\varphi_{\boldsymbol{r}}(t)/2}\psi_{\boldsymbol{r}\alpha,2}, 
\end{split}
\label{Eq:GaugeTransformation}
\end{equation}
Focusing on the temporal fluctuation in the phase $\varphi_{\boldsymbol{r}}(t)$, the Lagrangian changes to
\begin{equation}
\begin{split}
    \mathcal{L} = &\bar{\psi}\left[i\partial_t-H_0\right]\psi\\
    -&\frac{1}{2}\dot{\varphi}_{\boldsymbol{r}}(t)\left(\sum_{\alpha}\bar{\psi}_{\alpha}\sigma_z\psi_{\alpha}\right) - \frac{V}{2}\left(\sum_{\alpha}\bar{\psi}_{\alpha}\sigma_z\psi_{\alpha}\right)^2\\
    +&\sum_{\alpha}\Delta_{\boldsymbol{r},\alpha}\bar{\psi}_{2,\alpha}(\boldsymbol{r})\psi_{1,\alpha}(\boldsymbol{r})+\bar{\Delta}_{\boldsymbol{r},\alpha}\bar{\psi}_{1,\alpha}(\boldsymbol{r})\psi_{2,\alpha}(\boldsymbol{r})
\end{split}
\end{equation}
This is a Lagrangian of interacting fermions. Simplifications can be made by introducing additional auxiliary fields.

Namely, we introduce two auxiliary fields $\theta_{\boldsymbol{r}}$ and $\lambda_{\boldsymbol{r}}$ to the Lagrangian: 
\begin{equation}
    \delta\mathcal{L}=-\lambda_{\boldsymbol{r}}\left(\theta_{\boldsymbol{r}}-\sum_{\alpha}\bar{\psi}_{\alpha}\sigma_z\psi_{\alpha} \right),
\end{equation}
This term corresponds to adding an identity to the path integral. Indeed, integrating out $\lambda_{\boldsymbol{r}}$, we obtain a $\delta$-function which identifies $\theta_{\boldsymbol{r}}=\sum_{\alpha}\bar{\psi}_{\alpha}\sigma_z\psi_{\alpha}$. Then, integrating out $\theta_{\boldsymbol{r}}$ gives identity. Hence, path integral is not changed by introducing the two auxiliary fields $\theta_{\boldsymbol{r}}$ and $\lambda_{\boldsymbol{r}}$.

With the auxiliary fields introduced above, we can replace most of the operators $\sum_{\alpha}\bar{\psi}_{\alpha}\sigma_z\psi_{\alpha}$ by $\theta_{\boldsymbol{r}}$ fields. The resultant Lagrangian reads:
\begin{equation}
\begin{split}
    \mathcal{L} = &\bar{\psi}\left[i\partial_t-H_0\right]\psi+\lambda_{\boldsymbol{r}}\sum_{\alpha}\bar{\psi}_{\alpha}\sigma_z\psi_{\alpha}-\lambda_{\boldsymbol{r}}\theta_{\boldsymbol{r}}\\
    -&\frac{1}{2}\dot{\varphi}_{\boldsymbol{r}}(t)\theta_{\boldsymbol{r}} - \frac{V}{2}\theta_{\boldsymbol{r}}^2\\
    +&\sum_{\alpha}\Delta_{\boldsymbol{r},\alpha}\bar{\psi}_{2,\alpha}(\boldsymbol{r})\psi_{1,\alpha}(\boldsymbol{r})+\bar{\Delta}_{\boldsymbol{r},\alpha}\bar{\psi}_{1,\alpha}(\boldsymbol{r})\psi_{2,\alpha}(\boldsymbol{r})
\end{split}
\end{equation}
At this step, it is straightforward to integrate out the fermions. The action reads:
\begin{equation}
    \begin{split}
        \mathcal{S}=&-i\ \text{Tr}\ln\left[i\partial_t-H_{\text{MF}}+\lambda_{\boldsymbol{r}}\sigma_z\right]\\
        +&\int dt\sum_{\boldsymbol{r}}\left[-\frac{1}{2}\dot{\varphi}_{\boldsymbol{r}}(t)\theta_{\boldsymbol{r}}(t)-\frac{V}{2}\theta^2_{\boldsymbol{r}}(t)-\lambda_{\boldsymbol{r}}(t)\theta_{\boldsymbol{r}}(t)\right]
    \end{split}
\end{equation}
We should comment on the first line. We make following assumptions on the first line. First, we expand the first line to second order in $\lambda_{\boldsymbol{r}}$ and focus on the local terms:
\begin{equation}
\resizebox{0.95\linewidth}{!}{$\begin{split}
        \mathcal{S}=\int dt\sum_{\boldsymbol{r}}\left[-\frac{1}{2}\dot{\varphi}_{\boldsymbol{r}}(t)\theta_{\boldsymbol{r}}(t)-\frac{V}{2}\theta^2_{\boldsymbol{r}}(t)-\lambda_{\boldsymbol{r}}(t)\theta_{\boldsymbol{r}}(t)+\frac{1}{2E_0}\lambda^2_{\boldsymbol{r}}\right]
    \end{split}$}
\end{equation}
where $E_0$ is a dimensionful parameter. Then, integrate out the field $\lambda_{\boldsymbol{r}}$:
\begin{equation}
    \begin{split}
        \mathcal{S}=\int dt\sum_{\boldsymbol{r}}\left[-\frac{1}{2}\dot{\varphi}_{\boldsymbol{r}}(t)\theta_{\boldsymbol{r}}(t)-\frac{V+E_0}{2}\theta^2_{\boldsymbol{r}}(t)\right]
    \end{split}
\end{equation}
Lastly, we make a change of variable $\Pi_{\boldsymbol{r}}(t)=-\frac{1}{2}\theta_{\boldsymbol{r}}(t)$:
\begin{equation}
    \begin{split}
        \mathcal{S}=\int dt\sum_{\boldsymbol{r}}\left[\dot{\varphi}_{\boldsymbol{r}}(t)\Pi_{\boldsymbol{r}}(t)-\frac{4(V+E_0)}{2}\Pi^2_{\boldsymbol{r}}(t)\right]
    \end{split}
\end{equation}
To this end, it is clear that the repulsive interaction introduces competing fluctuation to the problem. Indeed, a strong repulsive interaction tends to pin the value of the canonical momentum $\Pi_{\boldsymbol{r}}$, which leads to a random phase $\varphi_{\boldsymbol{r}}$ and destroys the phase coherence.

A few more comments are due. First, we assume $E_0=0$. This assumption is consistent with the mean field study to begin with. Indeed, $E_0=0$ suppresses the fluctuations in the field $\lambda_{\boldsymbol{r}}$. Namely, we can neglect all the terms with $\lambda_{\boldsymbol{r}}$ fields in the previous paragraph.

Second, in the maintext, we \emph{assumed} the repulsive interaction is much weaker than the attractive interaction, $V\ll U$. Notice that the mean field order parameter is on the order of the attractive interaction $\Delta\sim U$ over a wide range of parameter choice. Hence, we \emph{assume} such weak repulsive interaction does not kill the mean field order parameter over a wide range of parameter choice.

Lastly, the analysis of this subsection should be combined with the phase stiffness (or its discretized version of Josephson junction array). Combining this and previous subsection, we obtain the action or Hamiltonian for the phase dynamics as summarized below.

\subsection{Quantum Hamiltonian for Josephson junction array}

The action describing the phase dynamics is the following: ($E_C=4V$)
\begin{equation}
    \begin{split}
        \mathcal{S}=&\int dt\sum_{\boldsymbol{r}}\left[\dot{\varphi}_{\boldsymbol{r}}(t)\Pi_{\boldsymbol{r}}(t)-\frac{1}{2}E_C\Pi^2_{\boldsymbol{r}}(t)\right]\\
        +&E_J\int dt\sum_{\boldsymbol{r}\neq\boldsymbol{r}'} f(\boldsymbol{r},\boldsymbol{r}')\cos(\varphi_{\boldsymbol{r}}-\varphi_{\boldsymbol{r}'})
    \end{split}
\end{equation}

Correspondingly, one can read out the quantum Hamiltonian for the low energy phase dynamics as:
\begin{equation}\label{eq:jj_ham}
    \begin{split}
    H_{\text{JJ}} = -\frac{1}{2}E_C\sum_{\boldsymbol{r}}\frac{\partial^2}{\partial\varphi^2_{\boldsymbol{r}}} - E_J\sum_{\boldsymbol{r}\neq\boldsymbol{r}'} f(\boldsymbol{r}-\boldsymbol{r}')\cos(\varphi_{\boldsymbol{r}}-\varphi_{\boldsymbol{r}'}).
    \end{split}
\end{equation}
where the canonical momentum is identified as $\Pi_{\boldsymbol{r}}=-i\frac{\partial}{\partial\varphi_{\boldsymbol{r}}}$. Below, we solve this Hamiltonian with a mean field type anaylsis. 

\subsection{Mean field critical point}
We study the mean field solution to the Hamiltonian Eq.~(\ref{eq:jj_ham}) and determine the critical point separating incoherent fluid and superfluid.

Due to the $U(1)$ symmetry ($\varphi_{\boldsymbol{r}}\rightarrow\varphi_{\boldsymbol{r}}+\alpha$), we can define the following mean field Hamiltonian is
\begin{equation}\label{eq:jj_ham_mf}
    \hat H_{MF} = -\frac{1}{2}E_C\frac{\partial^2}{\partial\varphi^2}-\tilde E_J\ev{\cos\varphi}\cos\varphi
\end{equation}
with (here, summation is only over $\boldsymbol{r}^{\prime}$)
\begin{equation}
    \tilde E_J=\sum_{\boldsymbol{r}',\boldsymbol{r}'\neq \boldsymbol{r}} E_J f(\boldsymbol{r},\boldsymbol{r}'),
\end{equation}
and $\ev{\cos\varphi}$ is the expectation value with respect to the mean field \emph{groundstate} wavefunction $\Psi(\varphi)$,
\begin{equation}\label{eq:jj_ham_self_cons}
    \ev{\cos\varphi}=\int_0^{2\pi}d\varphi \cos\varphi |\Psi(\varphi)|^2.
\end{equation}
\begin{equation}
    \langle \sin\varphi\rangle=\int_0^{2\pi}d\varphi \sin\varphi |\Psi(\varphi)|^2=0.
\end{equation}

Numerically, one can parameterize the wavefunction as:
\begin{equation}
    \Psi(\varphi)=\sum_{n=-N}^Na_n\frac{e^{in\varphi}}{\sqrt{2\pi}}
\end{equation}
with proper truncation $N$. Then, rewrite the mean field Hamiltonian in the basis of the harmonics and diagonalize the Hamiltonain. Evaluating the expectation value of the cosine potential determines the phase diagram reported in the maintext. Indeed, the critical point locates at $\tilde{E}_{\text{J}}=\frac{1}{2}E_C$.

\section{Josephson coupling: a Microscopic Derivation}
In the previous sections, we phenomenologically discretize the superfluid stiffness term on a lattice. As a result, we  obtain the model of Josephson junction array to describing the phase dynamics. It is natural to ask if we can obtain a similar Josephson coupling as Eq.~(\ref{eq:jj_term}) from a microscopic analysis. In this section, we derive a qualitatively analogous term, by computing the effective action within path integral formalism. 

To start with, we consider the order parameter $\Delta_{\boldsymbol{r}} = \Delta e^{i\varphi_{\boldsymbol{r}}}$ with spatial phase fluctuation $\varphi_{\boldsymbol{r}}$ around the mean-field solution Eq.~(\ref{Eq:MFSolution}). In the path integral, after integrating out fermionic fields, we obtain the effective action
\begin{equation}
\begin{split}
    i\mathcal{S}[\bar\Delta, \Delta] &= \Tr\ln \left[ G^{-1}_{0} + \begin{bmatrix}
	0 & -\bar{\Delta}_{\boldsymbol{r}} \\
	-\Delta_{\boldsymbol{r}} & 0 \\
	\end{bmatrix}\right]\\
 &+ i\int dt\sum_{\boldsymbol{r}}\sum_{\alpha=\text{A,B,C}}\frac{\left|\Delta_{\alpha}\right|^2}{U},\\
    G_{0}^{-1} &= i\partial_t-H_0
\end{split}
\end{equation}
where the trace includes both matrix trace and the convolution of space-time coordinates. We aim at describing the fluctuations to quadratic order in the order parameter $\Delta_{\boldsymbol{r}}$. To achieve this goal, we perform second derivative on the action $S$ in $\Delta_{\boldsymbol{r}}$. More details are as follows.

The second derivative is expressed as
\begin{equation}
\begin{split}
&\Pi_{\bar{\Delta}\Delta}(t,t^{\prime},\boldsymbol{r},\boldsymbol{r}') = \frac{\delta^2 S[\bar\Delta,\Delta]}{\delta\bar\Delta_{\boldsymbol{r}}(t)\delta\Delta_{\boldsymbol{r}'}(t^{\prime})} \bigg|_{\Delta_{\boldsymbol{r}}=\Delta_{\text{MF}}} \\
=& i\text{tr} \left[G_{\text{MF}}(t,\boldsymbol{r};t',\boldsymbol{r}')
\begin{bmatrix}
    0 & 0 \\
    1 & 0
\end{bmatrix}G_{\text{MF}}(t',\boldsymbol{r}';t,\boldsymbol{r})\begin{bmatrix}
    0 & 1 \\
    0 & 0
\end{bmatrix}
\right].
\end{split}
\end{equation}
The action at the quadratic order is given by:
\begin{equation}
\begin{split}
    \mathcal{S}^{(2)}=&\int dt\int dt^{\prime}\sum_{\boldsymbol{r},\boldsymbol{r}^{\prime}}\bar{\Delta}_{\boldsymbol{r}}(t)\Delta_{\boldsymbol{r}^{\prime}}(t^{\prime})\Pi_{\bar{\Delta}\Delta}(t,t^{\prime},\boldsymbol{r},\boldsymbol{r}')\\
    \approx&\int dt\sum_{\boldsymbol{r},\boldsymbol{r}^{\prime}}\bar{\Delta}_{\boldsymbol{r}}(t)\Delta_{\boldsymbol{r}^{\prime}}(t)\int dt^{\prime}\Pi_{\bar{\Delta}\Delta}(t,t^{\prime},\boldsymbol{r},\boldsymbol{r}')
\end{split}
\end{equation}
In the second line, we assume the order parameter varies slowly in time. Below, we focus on the following quantity:
\begin{equation}
    \Xi(t,\boldsymbol{r},\boldsymbol{r}^{\prime})=\int dt^{\prime}\Pi_{\bar{\Delta}\Delta}(t,t^{\prime},\boldsymbol{r},\boldsymbol{r}')
\end{equation}

To proceed, we perform the calculation in the band basis as in the previous sections. 
The contribution from dispersive bands is suppressed at the large band gap limit. Hence, contributions from the flat bands dominate. Therefore, the quantity $\Xi(t,\boldsymbol{r},\boldsymbol{r}')$ is well approximated by
\begin{equation}
    \begin{split}
        \Xi(t,\boldsymbol{r}, \boldsymbol{r}') &= i\int \frac{d\omega}{2\pi}\int \frac{d^2k}{(2\pi)^2} \int\frac{d^2k^{\prime}}{(2\pi)^2} e^{i(\boldsymbol{k}-\boldsymbol{k'})\cdot(\boldsymbol{r}-\boldsymbol{r}')} \\ 
        &\times\text{tr}\left[ G_{\rm f}(\omega,\boldsymbol{k}) 
        \begin{bmatrix}
        0 & 0 \\
        u_{\f,\boldsymbol{k}}^\dagger u_{\f,\boldsymbol{k}'} & 0
        \end{bmatrix}
    G_{\f}(\omega,\boldsymbol{k'}) 
    \begin{bmatrix}
    0 & u_{\f,\boldsymbol{k}'}^\dagger u_{\f,\boldsymbol{k}} \\
    0 & 0
    \end{bmatrix}\right]   
    \end{split}
\end{equation}
Direct calculation shows that the result is given by
\begin{equation}
\begin{split}
    \Xi(t,\boldsymbol{r},\boldsymbol{r}') &= \frac{2\mu^2+\Delta^2}{4\left[\mu^2+\Delta^2\right]^{3/2} } f(\boldsymbol r,\boldsymbol r'),  \\
   f(\boldsymbol{r},\boldsymbol{r}') &= \int\frac{d^2k}{(2\pi)^2}\int\frac{d^2k'}{(2\pi)^2} e^{i(\boldsymbol{k}-\boldsymbol{k}')\cdot(\boldsymbol{r}-\boldsymbol{r}')} \left|u_{\boldsymbol{k}}^\dagger u_{\boldsymbol{k}'}\right|^2.
\end{split}
\end{equation}
Notice that the exact spatial function $f(\boldsymbol{r},\boldsymbol{r}^{\prime})$ motivates us to write down the phenomenological Josephson coupling.

To this end, we are able to obtain the effective action
\begin{equation}
\begin{split}
    \mathcal{S}^{(2)} &= E_{\text{J, micro}}\sum_{\boldsymbol{r},\boldsymbol{r}'}\int dt f(\boldsymbol{r},\boldsymbol{r}')e^{i(\varphi_{\boldsymbol{r}} -\varphi_{\boldsymbol{r}'}) } \\
    &= E_{\text{J, micro}}\int dt\sum_{\boldsymbol{r},\boldsymbol{r}'} f(\boldsymbol{r},\boldsymbol{r}') \cos(\varphi_{\boldsymbol{r}} - \varphi_{\boldsymbol{r}'})
\end{split}
\end{equation}
where the second line follows from $f(\boldsymbol{r},\boldsymbol{r}^{\prime})=f(\boldsymbol{r}^{\prime},\boldsymbol{r})$. The Josephson coupling strength is now given by $E_{\text{J, micro}}=\Delta^2\frac{2\mu^2+\Delta^2}{{4\left[\mu^2+\Delta^2\right]^{3/2}}}$. The dependence on filling $\nu$ of $S$ can be deduced by employing the mean-field solution. The result is given by
\begin{equation}
    E_{{\rm J,micro}}/U = \nu(1-\nu)\left[(\nu-\frac{1}{2})^2+\frac{1}{4}\right].
    \label{Eq:EJPrime_Micro}
\end{equation}
Despite its more complicated form, the expression above has its maximum at $\nu=\frac{1}{2}$. $E_{\rm J,micro}$ vanishes when both layers approach integer filling, $\nu=0,1$. Meanwhile, the spatial profile of the Josephson coupling is also dictated by the function $f(\boldsymbol{r},\boldsymbol{r}^{\prime})$. Therefore, in the current more microscopic derivation, qualitative feature of the Josephson coupling is the same as what we obtained from a phenomenoligical approach. Admittedly, the two approaches have some quantitative difference in the prefactors, namely $E_J$ (Eq.~(\ref{Eq:EJ_Pheno})) and $E_{\rm J,micro}$ (Eq.~(\ref{Eq:EJPrime_Micro})). Hence, the same quantum phase transition between incoherent fluid and superfluid phases is expected also in the current microscopic description.

\section{Enhance Josephson coupling by add additional hopping}
\label{Sec:JJCoupling_AddHopping}

\begin{figure}
    \centering
    \includegraphics[width=0.75\linewidth]{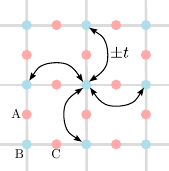}
    \caption{Additional hopping $\pm$ is added to the Lieb lattice model. As an example, the hopping between B-sites in adjacent unit cells is drawn explicitly. The same hopping amplitude is added to the A-sites and C-sites and is not drawn explicitly. Without loss of generality, the ``$-$'' sign is associated with layer 1 while ``$+$'' sign is for layer 2.}
    \label{Fig:AdditionalHopping}
\end{figure}

In this section, we show that adding \emph{additional} hopping terms enhance the Josephson coupling. As shown in Fig.~\ref{Fig:AdditionalHopping}, additional hopping terms can be added to Lieb lattice model between the same lattice sites in adjacent unit cells. Those additional hopping terms gives the flat band a dispersion $\epsilon(\boldsymbol{k})=\pm 2t\left(\cos k_x+\cos k_y\right)$. 

In terms of the exciton condensate, the additional hopping terms in Fig.~\ref{Fig:AdditionalHopping} provides an additional channel for exciton hopping (analog of pair hopping in the context of superconductivity). Thus, one would expect stronger phase synchronization effect, namely a stronger Josephson coupling.

\widetext
To demonstrate the point above, we consider the following path integral:
\begin{equation}
    \begin{split}
        \mathcal{Z}\left[\bar{\Delta},\Delta\right]=\int D\bar{\psi}D\psi e^{i\int dt\sum_{\boldsymbol{r},\boldsymbol{r}^{\prime}}\bar{\psi}(\boldsymbol{r})\left[G^{-1}_{\text{MF}}(t;\boldsymbol{r},\boldsymbol{r}^{\prime})\right]\psi(\boldsymbol{r}^{\prime})-i\int dt H_{\text{T}}-i\int dt H\left[\bar{\Delta},\Delta\right]}
    \end{split}
\end{equation}
where the last term $H\left[\bar{\Delta},\Delta\right]$ stems from the Hubbard-Stratonovich transformation and is not so important for the current discussion. Below, we will not write the last term explicitly. 

The additional hoppings are described by the Hamiltonian:
\begin{equation}
    H_{\text{T}}=\sum_{\boldsymbol{r}}\sum_{\delta\boldsymbol{r}}\sum_{\alpha}\left[-te^{i(\varphi_{\boldsymbol{r}}-\varphi_{\boldsymbol{r}+\delta\boldsymbol{r}})/2}\bar{\psi}_{1\alpha}(\boldsymbol{r})\psi_{1\alpha}(\boldsymbol{r}+\delta\boldsymbol{r})+te^{-i(\varphi_{\boldsymbol{r}}-\varphi_{\boldsymbol{r}+\delta\boldsymbol{r}})/2}\bar{\psi}_{2\alpha}(\boldsymbol{r})\psi_{2\alpha}(\boldsymbol{r}+\delta\boldsymbol{r})\right]
\end{equation}
where the phase factors $e^{\pm i(\varphi_{\boldsymbol{r}}-\varphi_{\boldsymbol{r}+\delta\boldsymbol{r}})/2}$ come from the gauge transformation, Eq.~(\ref{Eq:GaugeTransformation}). The summation is over nearest neighbor unit cells, namely $\delta\boldsymbol{r}=\pm\hat{x},\pm\hat{y}$. 

We treat the additional hopping terms as a perturbation and focus on the second order term:
\begin{equation}
    \mathcal{Z}^{(2)}=\int D\bar{\psi}D\psi\frac{1}{2}\left[-i\int dt H_{\text{T}}(t)\right]^2e^{i\int dt\sum_{\boldsymbol{r},\boldsymbol{r}^{\prime}}\bar{\psi}(\boldsymbol{r})\left[G^{-1}_{\text{MF}}(t;\boldsymbol{r},\boldsymbol{r}^{\prime})\right]\psi(\boldsymbol{r}^{\prime})}
\end{equation}
Among the various terms generated at the second order, there is a particular term describing the exciton hopping:
\begin{equation}
    \mathcal{Z}^{(2)}\approx-t^2\int dt_1\int dt_2\sum_{\boldsymbol{r},\delta\boldsymbol{r}}\sum_{\alpha\beta}e^{-i\left(\varphi(\boldsymbol{r},t_1)+\varphi(\boldsymbol{r},t_2)\right)/2}e^{i\left(\varphi(\boldsymbol{r}+\delta\boldsymbol{r},t_1)+\varphi(\boldsymbol{r}+\delta\boldsymbol{r},t_2)\right)/2}\langle \bar{\psi}_{1\alpha}(\boldsymbol{r},t_1) \psi_{2\beta}(\boldsymbol{r},t_2)\rangle \langle\bar{\psi}_{2\beta}(\boldsymbol{r}+\delta\boldsymbol{r},t_2)\psi_{1\alpha}(\boldsymbol{r}+\delta\boldsymbol{r},t_1)\rangle
\end{equation}
Then, assuming the phases $\varphi(\boldsymbol{r},t)$ varies slowly in time, the expression above can be further simplified:
\begin{equation}
    \mathcal{Z}^{(2)}\approx-t^2\int dt_1\int dt_2\sum_{\boldsymbol{r},\delta\boldsymbol{r}}\sum_{\alpha\beta}e^{i\left(\varphi(\boldsymbol{r}+\delta\boldsymbol{r},t_1)-\varphi(\boldsymbol{r},t_1)\right)}\langle \bar{\psi}_{1\alpha}(\boldsymbol{r},t_1) \psi_{2\beta}(\boldsymbol{r},t_2)\rangle \langle\bar{\psi}_{2\beta}(\boldsymbol{r}+\delta\boldsymbol{r},t_2)\psi_{1\alpha}(\boldsymbol{r}+\delta\boldsymbol{r},t_1)\rangle
\end{equation}
To this end, we neglected many terms, which involve spatially non-local single fermion Green's function. Indeed, when quantum metric is small, the single fermion Green's function is quite localized. Hence, we neglected the spatially non-local terms.

The next step is to utilize the Green's function: (the layer indices take the value of $l=1(2)$ and $\bar{l}=2(1)$)
\begin{equation}
    \begin{split}
        \langle\bar{\psi}_{l\alpha}(\boldsymbol{r},t_1) \psi_{\bar{l}\beta}(\boldsymbol{r},t_2)\rangle=&-i\left[G_{\text{MF}}\left(\boldsymbol{r},t_2;\boldsymbol{r},t_1\right)\right]_{\bar{l}\beta,l\alpha}\\
        =&-i\int\frac{d^2k}{(2\pi)^2}\frac{\Delta}{2\sqrt{\mu^2+\Delta^2}}\left[u_{\boldsymbol{k}}u^{\dagger}_{\boldsymbol{k}}\right]_{\beta\alpha}\\
        &\times\int\frac{d\omega}{2\pi}e^{-i\omega(t_1-t_2)}\left[\frac{1}{\omega+i0^+\text{sgn}(\omega)-\sqrt{\mu^2+\Delta^2}}-\frac{1}{\omega+i0^+\text{sgn}(\omega)+\sqrt{\mu^2+\Delta^2}}\right]
    \end{split}
\end{equation}
Direct calculation shows that:
\begin{equation}
    \begin{split}
        \mathcal{Z}^{(2)}\approx &i\frac{t^2\Delta^2}{4\left(\mu^2+\Delta^2\right)^{\frac{3}{2}}}\int\frac{d^2k}{(2\pi)^2}\int\frac{d^2k^{\prime}}{(2\pi)^2}\left|u^{\dagger}_{\boldsymbol{k}}u_{\boldsymbol{k}^{\prime}}\right|^2\int dt\sum_{\boldsymbol{r},\delta\boldsymbol{r}}e^{i\left(\varphi(\boldsymbol{r}+\delta\boldsymbol{r},t_1)-\varphi(\boldsymbol{r},t_1)\right)}\\
        =&i\frac{t^2\Delta^2}{4\left(\mu^2+\Delta^2\right)^{\frac{3}{2}}}\int\frac{d^2k}{(2\pi)^2}\int\frac{d^2k^{\prime}}{(2\pi)^2}\left|u^{\dagger}_{\boldsymbol{k}}u_{\boldsymbol{k}^{\prime}}\right|^2\int dt\sum_{\boldsymbol{r},\delta\boldsymbol{r}}\cos\left[\varphi(\boldsymbol{r}+\delta\boldsymbol{r},t_1)-\varphi(\boldsymbol{r},t_1)\right]
    \end{split}
\end{equation}
Therefore, the action for the phase fluctuations $\varphi(\boldsymbol{r},t)$ acquires a correction:
\begin{equation}
    \delta\mathcal{S}^{(2)}=\frac{t^2\Delta^2}{4\left(\mu^2+\Delta^2\right)^{\frac{3}{2}}}\int\frac{d^2k}{(2\pi)^2}\int\frac{d^2k^{\prime}}{(2\pi)^2}\left|u^{\dagger}_{\boldsymbol{k}}u_{\boldsymbol{k}^{\prime}}\right|^2\int dt\sum_{\boldsymbol{r},\delta\boldsymbol{r}}\cos\left[\varphi(\boldsymbol{r}+\delta\boldsymbol{r},t_1)-\varphi(\boldsymbol{r},t_1)\right]
\end{equation}
In turn, the correction in the action translates to a correction of Hamiltonian as:
\begin{equation}
    \delta H^{(2)}=-\frac{t^2\Delta^2}{4\left(\mu^2+\Delta^2\right)^{\frac{3}{2}}}\int\frac{d^2k}{(2\pi)^2}\int\frac{d^2k^{\prime}}{(2\pi)^2}\left|u^{\dagger}_{\boldsymbol{k}}u_{\boldsymbol{k}^{\prime}}\right|^2\sum_{\boldsymbol{r},\delta\boldsymbol{r}}\cos\left[\varphi(\boldsymbol{r}+\delta\boldsymbol{r},t_1)-\varphi(\boldsymbol{r},t_1)\right]
\end{equation}
Notice the minus sign in the front. Indeed, the additional hoppings enhances the Josephson coupling.


\bibliography{ExcitonCondensate}